\documentclass[preprint]{aastex}

\usepackage{times}

\newcommand\lath{\hbox{$^{\textrm h}$}}
\newcommand\latm{\hbox{$^{\textrm m}$}}
\newcommand\lats{\hbox{$^{\textrm s}$}}

\newcommand\mum{$\mu$m}
\newcommand\nhp{N$_2$H$^+$}
\newcommand\kms{km\,s$^{-1}$}
\newcommand\kmspc{km\,s$^{-1}$\,pc$^{-1}$}

\title{BIMA N$_2$H$^+$ 1-0 mapping observations of L183 -- fragmentation and spin-up in a collapsing, magnetized, rotating, pre-stellar core}
\shorttitle{Fragmentation and spin-up in a collapsing, rotating, pre-stellar core}
\shortauthors{Kirk, Crutcher, and Ward-Thompson}

\author{Jason M. Kirk \altaffilmark{1}}
\affil{School of Physics and Astronomy, Cardiff University, Queens Buildings, The Parade, Cardiff, CF24 3AA, United Kingdom}
\email{jason.kirk@astro.cf.ac.uk}
\and

\author{Richard M. Crutcher}
\affil{Department of Astronomy, University of Illinois at Urbana-Champaign, 1002 West Green Street, Urbana, Il 61801}
\email{crutcher@uiuc.edu}
\and

\author{Derek Ward-Thompson}
\affil{School of Physics and Astronomy, Cardiff University, Queens Buildings, The Parade, Cardiff, CF24 3AA, United Kingdom}
\email{derek.ward-thompson@astro.cf.ac.uk}

\altaffiltext{1}{Department of Astronomy, University of Illinois at Urbana-Champaign, 1002 West Green Street, Urbana, Il 61801}

\begin{document}


\begin{abstract}
We have used the Berkeley-Illinois-Maryland Array (BIMA) to make deep \nhp\ 1-0 maps of the pre-stellar core L183, in order to study the spatial and kinematic substructure within the densest part of the core. Three spatially and kinematically distinct clumps are detected, which we label L183-N1, L183-N2 and L183-N3. L183-N2 is approximately coincident with the submillimetre dust peak and lies at the systemic velocity of L183. Thus we conclude that L183-N2 is the central dense core of L183. L183-N1 and 3 are newly-discovered fragments of L183, which are marked by velocity gradients that are parallel to, but far stronger than, the velocity gradient of L183 as a whole, as detected in previous single-dish data. Furthermore, the ratio of the large-scale and small-scale velocity gradients, and the ratio of their respective size-scales, are consistent with the conservation of angular momentum for a rotating, collapsing core undergoing spin-up. The inferred axis of rotation is parallel to the magnetic field direction, which is offset from its long axis, as we have seen in other pre-stellar cores. Therefore, we propose that we have detected a fragmenting, collapsing, filamentary, pre-stellar core, rotating about its B-field, which is spinning up as it collapses. It will presumably go on to form a multiple protostellar system.
\end{abstract}



\section{Introduction}

The star formation process begins within molecular clouds, where the internal pressures (magnetic, turbulent, thermal, or otherwise), supporting dense regions against collapse, lose out to gravity. The densest regions are known as pre-stellar cores \citep{1994wsha}, which are the gravitationally bound cores of clouds that are believed to be undergoing star formation, but as yet show no signature of an embedded protostellar source \citep{2007wacjow}. Recent studies have revealed the presence of extremely faint protostellar objects inside a small number of cores that were previously believed to be pre-stellar \citep{2005crapsivello,2006bourke}. These appear to be very young protostars embedded deep within these cores.

However, the manner of the collapse from a pre-stellar core to a core that contains a protostar is still a matter of some debate -- for a review, see \citet{wtsci}. Gravitational collapse, rotation, turbulent motions and fragmentation may all play a role. Detailed study of the densest pre-stellar cores that do not contain a protostar is necessary to understand the evolution of pre-stellar cores \citep{jwt}.

The centrally condensed, high column density pre-stellar core L183, which is also known as L134N, is part of a loose group of high latitude ($b\sim$37\degr) dense molecular cores associated with the surface of the local bubble at a distance of $110\pm10$ pc \citep{1989franco}. It has been extensively studied with a number of different spectroscopic species \citep{2001lee,2000dickens,2005pagani,2007torres,2007pagani} and in the far-infrared and submillimetre continuum \citep{2002juvela, 2002wak, 2003lehtinen, 2003pagani, 2005kwa, 2008kauffmann}. It also appears to be more chemically evolved than other pre-stellar cores \citep{2000dickens} and displays a number of signatures common to cores that are believed to be close to forming a protostar \citep{2001lee, 2005crapsi}. Thus, L183 is a good candidate in which to study pre-stellar evolution.

In this paper we present new \nhp\ observations performed with the BIMA interferometer. \nhp\ is one of the preferred tracers for cold dense molecular material as it is well correlated with submillimetre dust emission and is less prone to gas phase depletion than other molecular species within pre-stellar cores \citep{jwt, 2002bergin, 2004tafalla}, and it has been widely used to survey pre-stellar cores in general and L183 in particular \citep{2000dickens,2002caselli,2005crapsi,2007pagani}. 

In Section \ref{obs} we describe the configuration of the BIMA array and in Section \ref{results} we show channel and integrated maps of \nhp\ intensity. In Section \ref{analysis} we show the results of fitting various line parameters, and we discuss their implications in Section \ref{discuss}.  


\section{Observations}
\label{obs}
BIMA\footnote{Operated by the Berkeley-Illinois-Maryland Association with NSF and University Funding. The BIMA facilities have since been merged with the OVRO array to form the present CARMA array.} was used to map the 3.22~mm (93~GHz) $J = 1\rightarrow0$ transition of \nhp\ \citep{1995caselli,2006daniel} towards the peak of the L183 submillimetre dust emission\footnote{$\alpha_{2000}$=15\lath 54\latm 08.8\lats, $\delta_{2000}$ $-$02\degr 52\arcmin 38\arcsec \citep{2005kwa}.}. BIMA was a 10-element millimeter-wave interferometer operated at the Hat Creek Radio Observatory in Northern California \citep{1996welch}. 

BIMA was used in its compact C-array ($\sim$2--27 k$\lambda$) and ultra-compact D-array ($\sim$2--9 k$\lambda$) configurations. Table \ref{tab:obs} shows the individual observation dates, durations and configurations. A total duration of 50 hours ($\sim$6 tracks) was spent in the C-Array configuration and a total duration of 18 hours ($\sim$2 tracks) was spent in the D-Array configuration. 

Our observations are only sensitive to angular scales between the FWHM of the primary beam (given by the effective diameter of an individual dish) and that of the synthesized beam (given by the uv sampling and weighting). The individual antennae have a physical diameter of 6.2~m and an effective primary beam FWHM of 1.7~arcmin at 3.22~mm. The synthesized-beam FWHM for the combined observations was $13\times10$ arcsec with a semi-major axis position angle of $-$14\degr. The beam was elongated due to the declination of the source. Therefore, these observations are sensitive to structure between 10 and 100 arcsec. This corresponds to a spatial scale of 0.005--0.05~pc at 110~pc.

\begin{table}
	\caption{Observation dates, duration and array configuration for the data used in this paper.}
	\label{tab:obs}
	\centering{
		\begin{tabular}{lcc}
			\hline
			Date & Duration & Configuration \\
			\hline
			2003 March 23 & 8 hrs & C-Array \\
			2003 March 24 & 8 hrs & C-Array \\
			2003 May 03 & 5 hrs & C-Array \\
			2003 May 07 & 8 hrs & C-Array \\
			2003 May 08 & 4 hrs & C-Array \\
			2003 May 15 & 5.5 hrs & C-Array \\
			2003 July 16 & 7 hrs & D-Array \\
			2003 July 17 & 7 hrs & D-Array \\
			2004 April 30 & 3 hrs & C-Array \\
			2004 May 09 & 8 hrs & C-Array \\
			2004 May 14 & 4 hrs & D-Array \\
			\hline
		\end{tabular}
	}
\end{table}

The BIMA correlator was set to mode 5 to give 512 channels across a bandwidth of 6.25~MHz. The channel width was 12.2~kHz (0.039 km~s$^{-1}$). The \nhp\ lineset was placed in the first sideband while the 96.4~MHz C$^{34}$S 2-1 transition was in the opposite sideband, but no signal was detected in the second sideband. Phase and amplitude calibration were conducted against the radio source 1549+026. The assumed flux of 1549+026 at our wavelength was 1.5 Jy (BIMA Calibrator List\footnote{http://bima.astro.umd.edu/cgi-bin/swflux\_plot.pl?source=1549\%2B026}).  The data were reduced using the MIRIAD software package and the visibilities were Fourier transformed into the image plane using natural weighting in order to maximize the sensitivity \citep{1995sault}. The theoretical per channel rms was 0.121 K.


\section{Results}
\label{results}

\begin{figure}
	\centering{
		\includegraphics[angle=270,width=0.8\columnwidth]{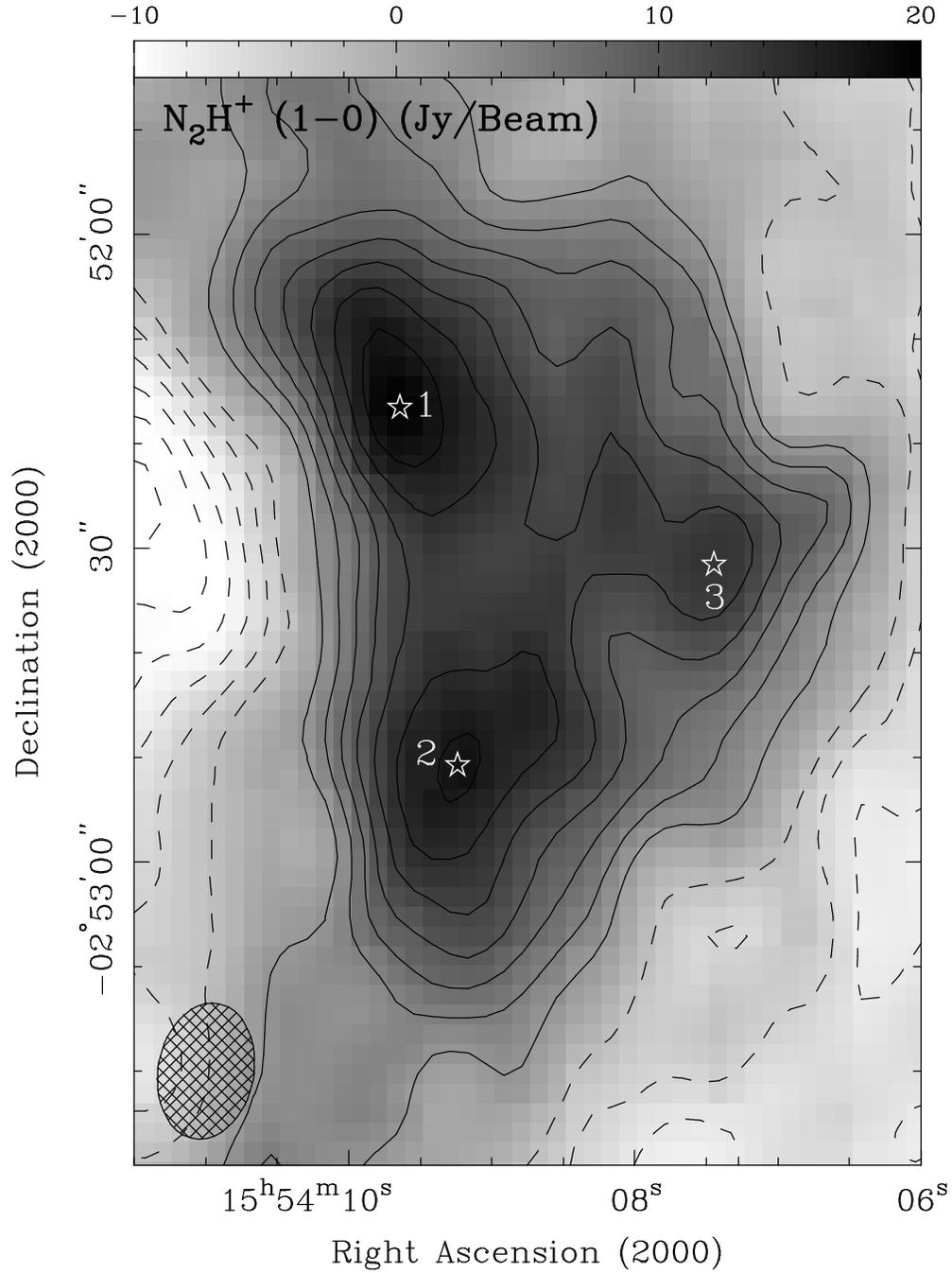}
		}
	\caption{A map of \nhp\ (1-0) line emission integrated across the seven hyperfine lines towards the L183 core. The greyscale shows the full range of emission, the contours start at, and have intervals of 2.46 Jy/Beam. The dashed contours are negative, they start at, and have intervals of $-$2.46 Jy/Beam. The hatched ellipse shows the size of the synthesized beam. The three sources, L183-N1, 2 \& 3, are marked with stars and are numbered. The contour spacing is 1.5 times the off-source rms of 1.64 Jy/Beam.}
	\label{fig:intmap}
\end{figure}

The total intensity of emission from the seven hyperfine components of \nhp\ towards the L183 core is shown in Figure~\ref{fig:intmap}, integrated over velocity. The first solid contour is at 2.46 Jy/Beam, which is also the separation between subsequent contours. For this dataset the scaling between Jy/Beam and Kelvin is 1.07 K/(Jy/Beam).

\begin{table*}
	\caption{Table of positions and hyperfine fit parameters towards the three peaks in the integrated \nhp\ map (Figure \ref{fig:intmap}).
		Column 1 lists the source name. 		
		Columns 2 and 3 list the central position of each source. 
		Columns 4 to 7 list the CLASS hfs parameters -- radial velocity ($V_{lsr}$), linewidth ($\Delta V$), total optical depth summed across all seven hyperfine components ($\tau$), and the main beam temperature ($T_{MB}$) -- for a fit to the spectra at the positions listed in columns 2 and 3. 
		Column 8 lists the non-thermal linewidth ($\Delta V_{NT}$) based on the assumption that $T_{K}=7$~K (see Section \ref{linewidths}). 
		Column 9 lists the rms of the radial velocity ($V_{rms}$) computed in a 5 by 5 pixel box. 
		Column 10 lists the excitation temperature ($T_{ex}$).
		Columns 11 and 12 list the magnitude and direction, and their standard deviations, of the local velocity gradient (see Section \ref{discuss}).
		}
	\label{tab:hfs}

	\centering{\resizebox{\textwidth}{!}{

		
			
		\begin{tabular}{
			l 					
			r@{\lath}c@{\latm}l@{\lats\hspace{0.5cm}}	
			r@{\degr}c@{\arcmin}l@{\arcsec\hspace{0.5cm}} 	
			r@{$\pm$}l 				
			r@{$\pm$}l 				
			r@{$\pm$}l 				
			r@{$\pm$}l 				
			c 					
			c 					
			r@{$\pm$}l 				
			r@{$\pm$}l				
			r@{$\pm$}l }				

			
			\hline	
			Source & 				
			\multicolumn{3}{c}{R.A.} & 		
			\multicolumn{3}{c}{Dec.} & 		
			\multicolumn{2}{c}{$V_{lsr}$} & 	
			\multicolumn{2}{c}{$\Delta V$} & 	
			\multicolumn{2}{c}{$\tau$} & 		
			\multicolumn{2}{c}{$T_{MB}$} & 		
			$\Delta V_{NT}$ &			
			$V_{rms}$ & 				
			\multicolumn{2}{c}{$T_{ex}$} & 		
			\multicolumn{2}{c}{$\mathcal{G}$} & 	
			\multicolumn{2}{c}{PA$_{\mathcal{G}}$}	
			\\
			
			&					
			\multicolumn{3}{c}{(2000)} & 		
			\multicolumn{3}{c}{(2000)} & 		
			\multicolumn{2}{c}{[km\,s$^{-1}$]} & 	
			\multicolumn{2}{c}{[km\,s$^{-1}$]} & 	
			\multicolumn{2}{c}{ } & 		
			\multicolumn{2}{c}{[K]} & 		
			[\kms] & 				
			[\kms] & 				
			\multicolumn{2}{c}{[K]} & 		
			\multicolumn{2}{c}{[\kmspc]} &		
			\multicolumn{2}{c}{[\degr]} \\		
		
			\hline

			
			L183-N1 & 					
			15 & 54 & 09.5 & 			
			$-$02 & 52 & 15 & 			
			2.256 & 0.003 & 			
			0.123 & 0.006 & 			
			0.37 & 0.08 & 				
			1.2 & 0.4 & 				
			0.064 & 				
			0.054 & 				
			6.5 & 2.0 & 				
			6.5 & 5.8 & 				
			$-$74 & 23 \\				
			
			L183-N2 & 					
			15 & 54 & 09.1 & 			
			$-$02 & 52 & 50 & 			
			2.415 & 0.004 & 			
			0.127 & 0.008 & 			
			2.1 & 0.6 & 				
			0.5 & 0.2 & 				
			0.071 & 				
			0.025 & 				
			3.3 & 1.8 & 				
			1.0 & 2.4 & 				
			\multicolumn{2}{c}{---} \\		
			
			L183-N3 &					
			15 & 54 & 07.3 & 			
			$-$02 & 52 & 31 & 			
			2.588 & 0.005 & 			
			0.108 & 0.010 & 			
			0.36 & 0.18 & 				
			0.63 & 0.45 & 				
			0.025 & 				
			0.090 & 				
			4.8 & 3.8 & 				
			8.9 & 8.0 & 				
			$-$57 & 60 \\				
			
			\hline
		\end{tabular}
	} }
\end{table*}

The integrated emission shown in Figure \ref{fig:intmap} displays the characteristic north-south L183 filament as seen in submillimetre dust emission, infra-red extinction, and single dish line maps of \nhp\ 1-0 \citep[e.g.][]{2003pagani,2005kwa,2005crapsi}. The integrated \nhp\ intensity shows three bright peaks (north, south and west). For reference we will refer to these features as L183 \nhp\ (L183-N) 1, 2, and 3 in order of their peak intensity. Their positions are listed in columns 2 and 3 of Table \ref{tab:hfs} and are shown in Figure \ref{fig:intmap} and in subsequent figures as labeled open stars.

\begin{figure}
	\centering{
	\includegraphics[angle=270,width=0.8\columnwidth]{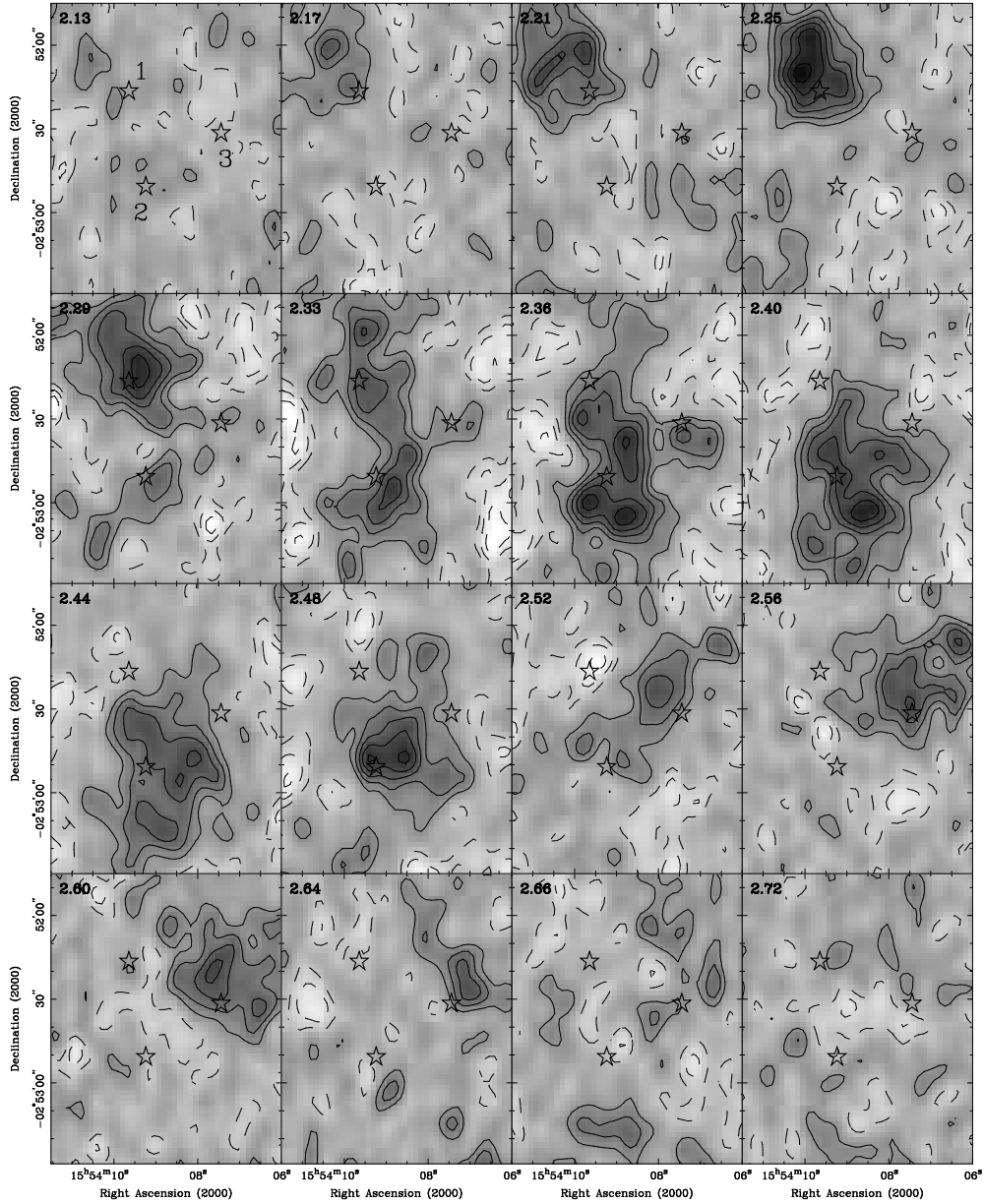}
	}
	\caption{Channel map across 2.15--2.72 km~s$^{-1}$ of the \nhp\ 101-012 hyperfine component towards L183. The channel width is 0.039 \kms. The velocity of each channel is shown in the top-left corner of each panel. The contours are in increments of 0.156 Jy/Beam (0.169~K) with the dashed contours mirroring negative values of the positive solid contours. The stars are as in Figure \ref{fig:intmap}, but only the stars in the 2.13~\kms\ panel are labeled. The contour spacing is 1.5 times the off-source rms of 0.104 Jy/Beam.}
	\label{fig:linegrid}
\end{figure}

To examine the kinematic structure of L183 we extracted channel maps of \nhp\ intensity across the 101-012 component as 0.039 \kms\ slices between 2.15 and 2.72 \kms. This component is shown as it is isolated in velocity from the other components and thus is free from blending between components. These channel maps are shown in Figure~\ref{fig:linegrid}. As the velocity increases, the emission first appears in the northeast, before moving to the south, and finishing in the northwest. These components roughly correspond to the three emission peaks seen in Figure~\ref{fig:intmap}. Thus, L183-N1 and L183-N3 are separated spatially and kinematically from each other, but are both connected kinematically to L183-N2.

\begin{figure}
	\centering{
		\includegraphics[angle=270,width=0.7\columnwidth]{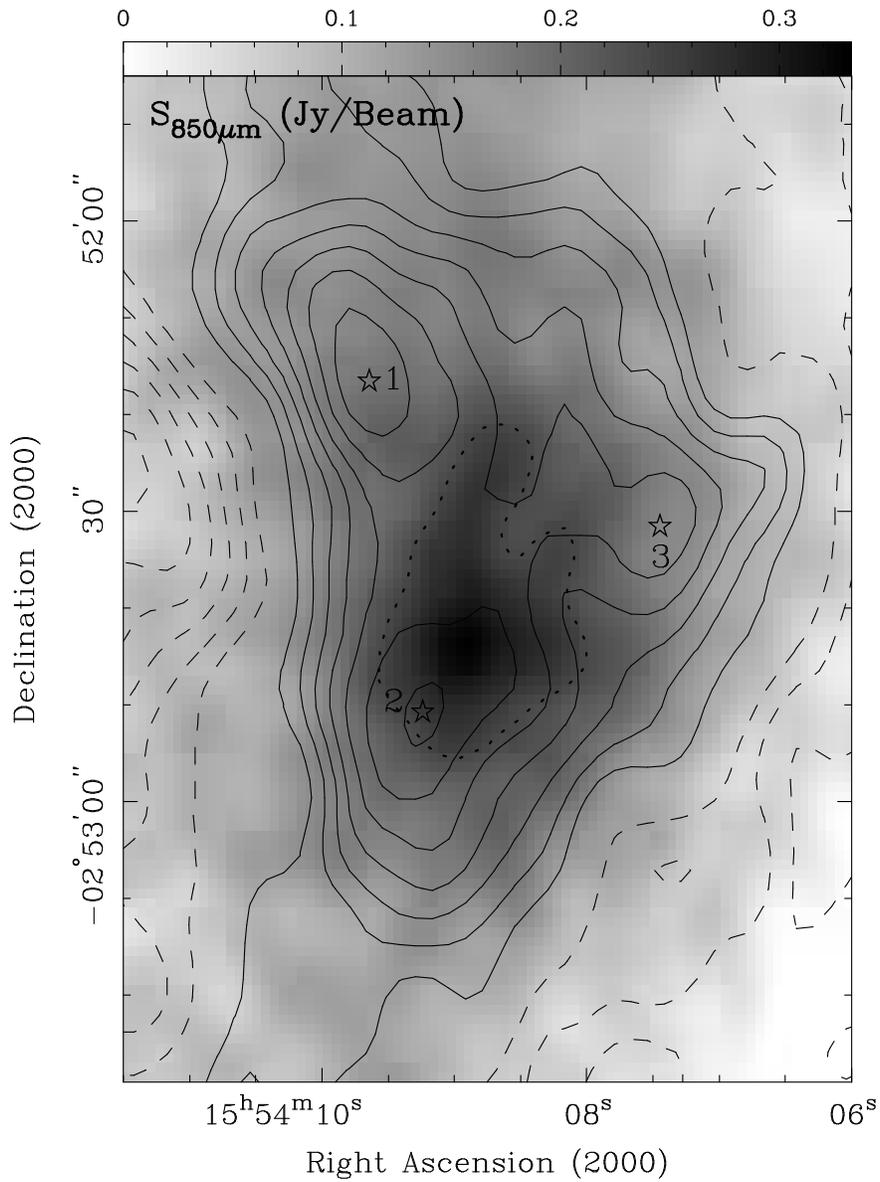}
	}
	\caption{Integrated \nhp\ contours from Figure~\ref{fig:intmap} overlaid on a greyscale image of SCUBA 850-\micron\ dust emission from \citet{2005kwa}. The dotted-contour is at 0.16 Jy/Beam and traces the central part of the SCUBA emission. The stars and labels are as in Figure \ref{fig:intmap}.} 
	\label{fig:scuba}
\end{figure}

The central section of the SCUBA 850-\micron\ map from \citet{2005kwa} is shown as a greyscale image in Figure~\ref{fig:scuba}. The SCUBA continuum emission peaks close to L183-N2 and has a northward extension between L183-N1 and 3. This extension is seen to continue on a larger scale in the SCUBA image \citep{2005kwa}. The L183-N2 peak is not exactly coincident with the SCUBA 850-\mum\ emission peak. The peak in the continuum emission is in fact coincident with the \nhp\ emission at 2.44--2.48~\kms.

\section{Spectral Line Analysis}
\label{analysis}

The data shown in Figure \ref{fig:intmap} use a pixel scale of 1.87 arcsec per pixel in order to show a smooth intensity distribution. For the analysis of the spectral lines the data were spatially averaged onto a 5.6 arcsec pixel grid in order to produce pixels that are approximately Nyquist sampled and to generate a factor of 3 increase in signal-to-noise ratio. The spectrum for each pixel that had a 2$\sigma$ or greater feature at $\sim$2.5 \kms\ (the approximate location of the 101-012 hyperfine component) was then extracted and piped to the hyperfine structure (hfs) routine from the CLASS data reduction package\footnote{CLASS is part of the GILDAS software suite available from \url{http://www.iram.fr/IRAMFR/GILDAS}} \citep{hfsMan,classMan}. The hfs routine was used to simultaneously fit all seven hyperfine components giving best fits for the radial velocity $V_{lsr}$, the linewidth $\Delta V$, the total optical depth $\tau$ and the product of the main beam temperature and optical depth ($T_{MB}\times\tau$). For the fitting it was assumed that all hyperfine components had the same linewidth and excitation temperature. 

\begin{figure}
	\centering{
		\includegraphics[angle=270,width=0.5\columnwidth]{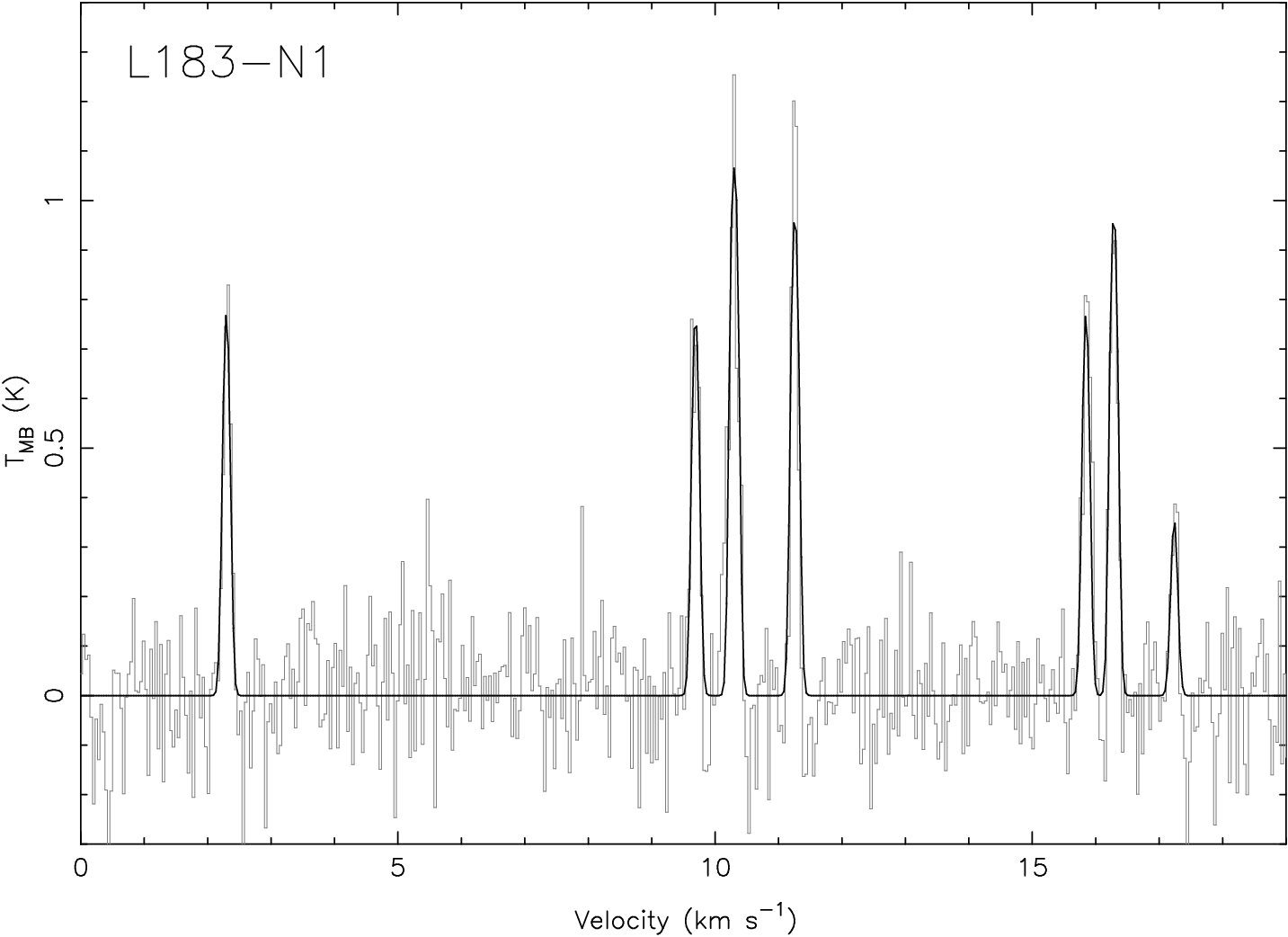} \\
		\includegraphics[angle=270,width=0.5\columnwidth]{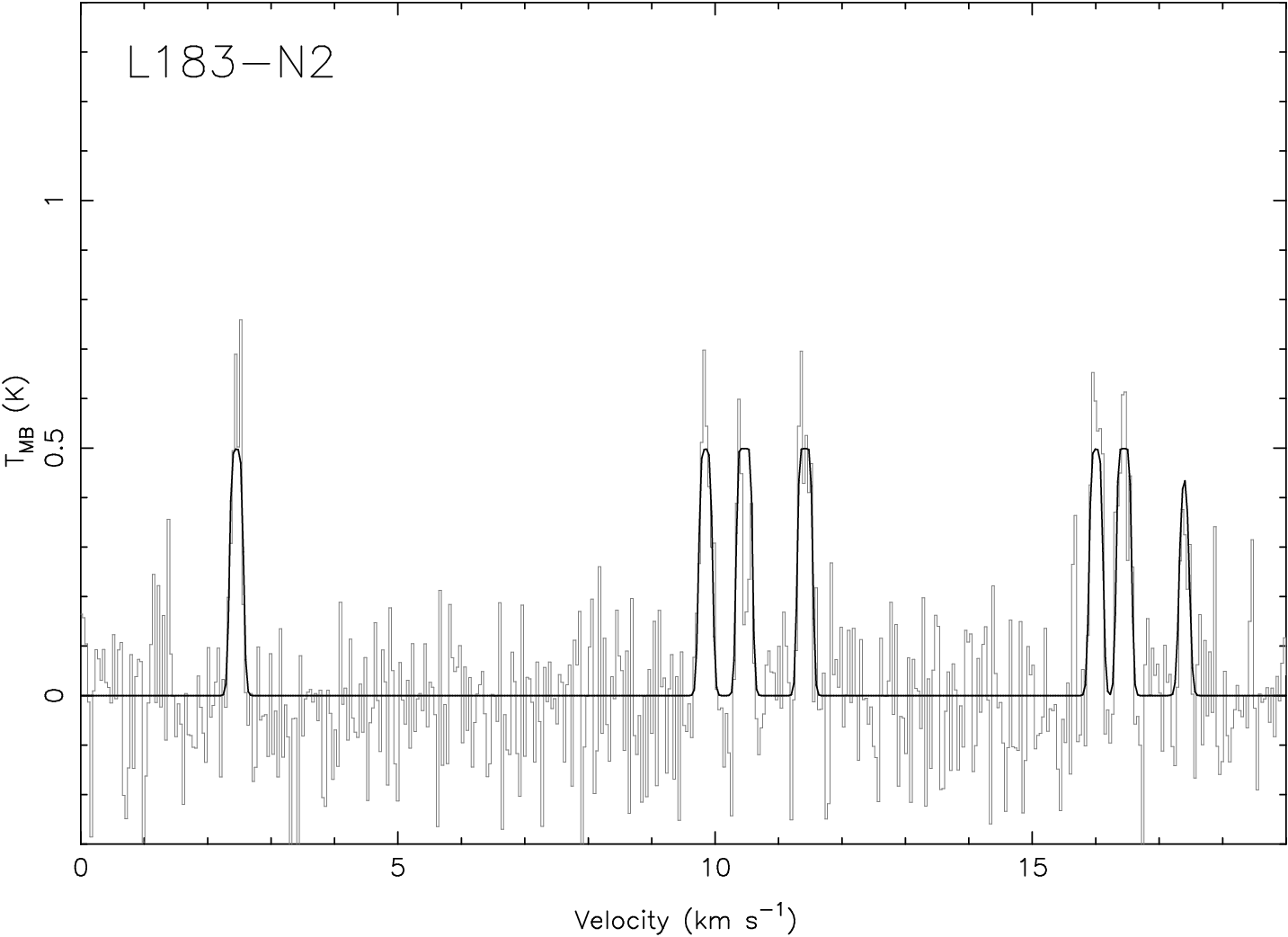} \\
		\includegraphics[angle=270,width=0.5\columnwidth]{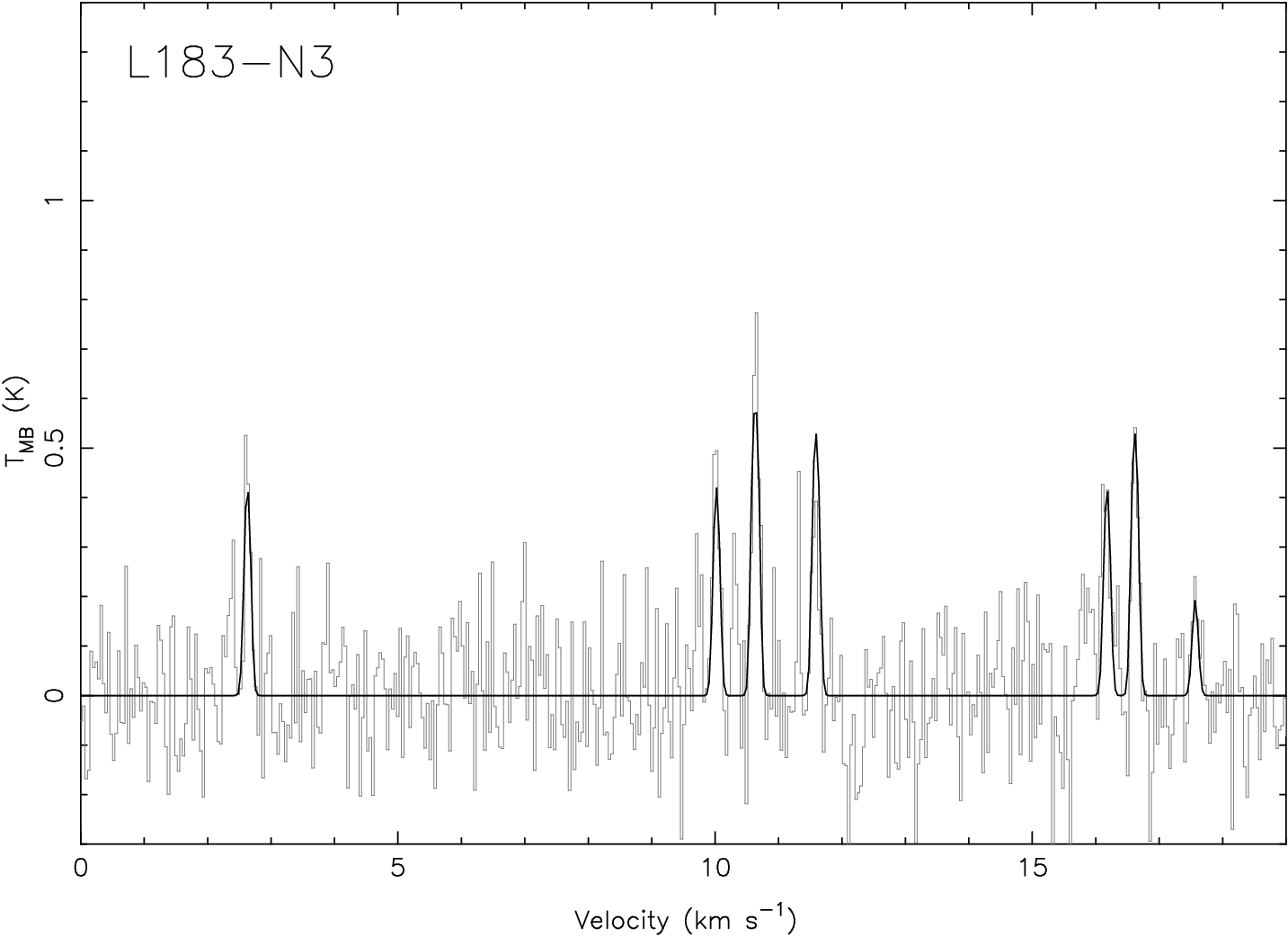} \\
	}
	\caption{\label{fig:specs}The \nhp\ spectrum (gray line) observed observed towards the L183-N1, 2, and 3 positions averaged over a 5.6 arcsec box. The black line is the CLASS hyperfine fit to each position. The parameters of each fit are listed in Table \ref{tab:hfs}.} 
	
\end{figure}

Table \ref{tab:hfs} lists the fit parameters at the positions of the three peaks identified from the map of integrated emission. The spectra for these positions are shown in Figure \ref{fig:specs}. We note that the errors on the fitting of $V_{lsr}$ and $\Delta_{V}$ are narrower than the channel width (0.039~\kms), due to multiple channels over the line that contribute to the fitting. 

\subsection{Kinematics}
\label{kinematics}

\begin{figure}
	\centering{
		\includegraphics[angle=270, width=0.8\columnwidth]{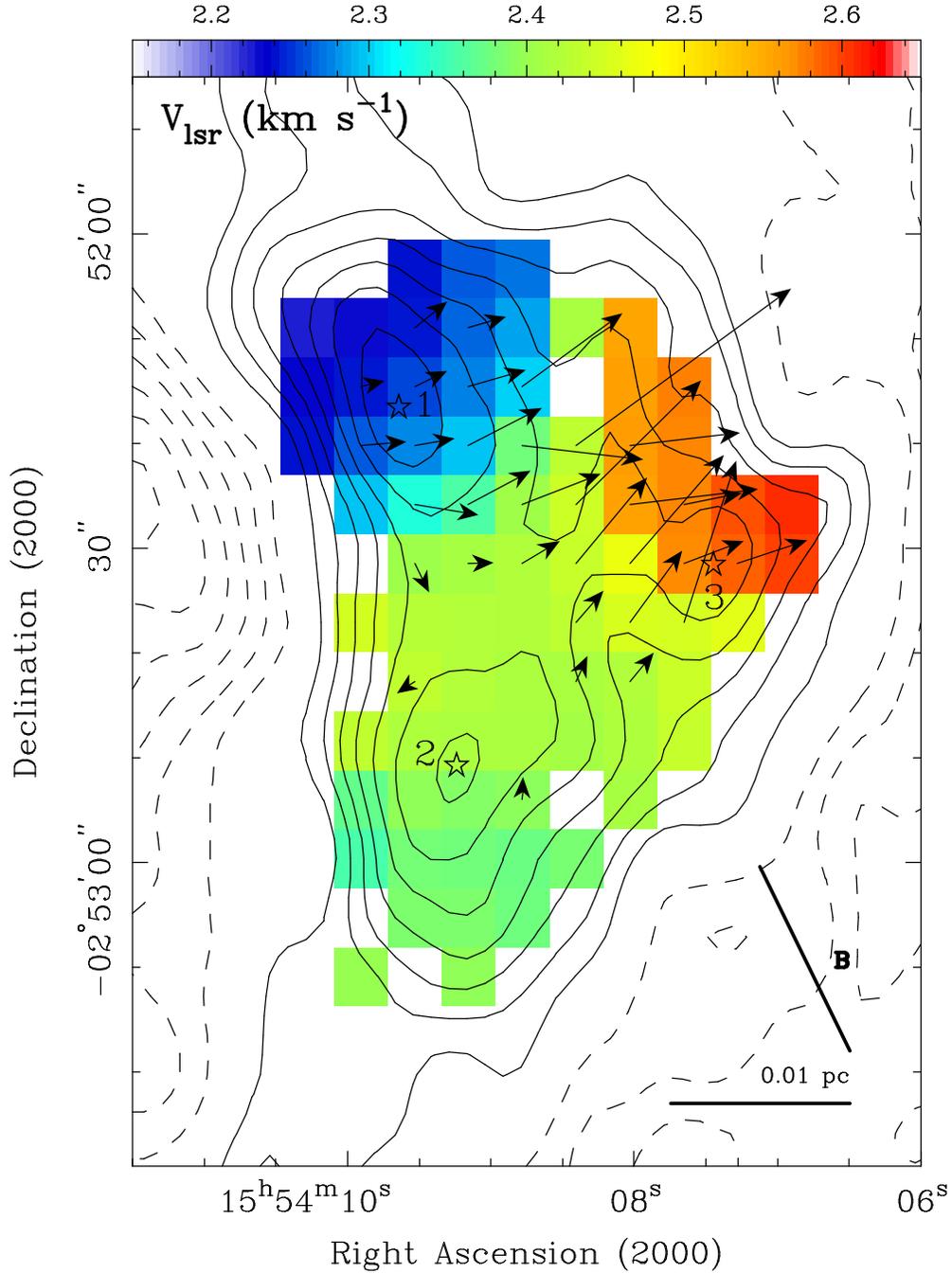}
	}
	\caption{A map of radial velocity. The overlaid contours are the integrated contours from Figure \ref{fig:intmap}. The range of the velocities is shown by the bar at the top of the plot. The three labeled stars show the positions listed in Table \ref{tab:hfs}. The plot is annotated with a 0.01 pc linear scale bar and a black-line showing the orientation of the plane-of-the-sky magnetic field as measured by \citet{2004crutcher}. The vectors show the local gradient of the $V_{lsr}$. The length of the vectors shows the size and direction of the velocity gradient with 1 arcsec equal to 1 \kmspc.} 
	\label{fig:vlsr}
\end{figure}

Figure \ref{fig:vlsr} shows the variation of the local standard of rest velocity ($V_{lsr}$) across the L183 molecular core. It shows the same triple-lobed structure as shown in Figure \ref{fig:linegrid}, but now we see that the most extreme velocities are towards the outside of each of the three lobes. The velocities at the position of each of the three L183-N sources are listed in column 4 of Table \ref{tab:hfs}. There is a 0.33~\kms\ separation between L183-N1 and L183-N3.

There have been previous observations of \nhp\ towards L183 using the single-dish IRAM 30-meter telescope. Its beamwidth at 3.22~mm is $\sim$30 arcsec. \citet{2007pagani} measured a value of $V_{lsr}$ = 2.367 km\,s$^{-1}$ towards RA(2000) = 15\lath 54\latm 08.5\lats, Dec.(2000) = $-$02\degr 52\arcmin 48\arcsec. This is in good agreement with the values we measure in that area. \citet{2005crapsi} measured a value of $V_{lsr}$ = 2.413 km\,s$^{-1}$ towards RA(2000) = 15\lath 54\latm 08.4\lats, Dec.(2000) = $-$02\degr 52\arcmin 23\arcsec. This again is similar to the values we measure exactly at that point. The literature value for the $V_{lsr}$ of the whole core of L183 (2.415 km~s$^{-1}$) matches closely that of L183-N2. Hence we take 2.415 km~s$^{-1}$ as the systemic velocity of the L183 core and we see that the velocities of L183-N1 and L183-N3 are velocity structures overlaid on this, of up to $\pm$0.17~kms$^{-1}$.

To examine the velocity gradient across L183 we averaged the gradient vectors between a single pixel and its nearest neighbors with the condition that each pixel must have at least seven neighbors with valid measurements of $V_{lsr}$. A vector map of the local velocity gradient is overlaid on the map of $V_{lsr}$ in Figure \ref{fig:vlsr}.

Figure \ref{fig:vlsr} also shows a black-line indicating the direction of the plane-of-the-sky magnetic field as observed by \citet{2004crutcher}. This is approximately orthogonal to the large-scale velocity gradient (see Section \ref{discuss}). The magnetic field may therefore play an important role in the evolution of this pre-stellar core \citep[c.f.][]{hgwa}. We discuss further the relevance of the velocity gradient to previous observations in Section \ref{discuss}.


\subsection{Linewidths}
\label{linewidths}

\begin{figure*}
	\centering{
	 
	\includegraphics[angle=270,width=0.24\textwidth]{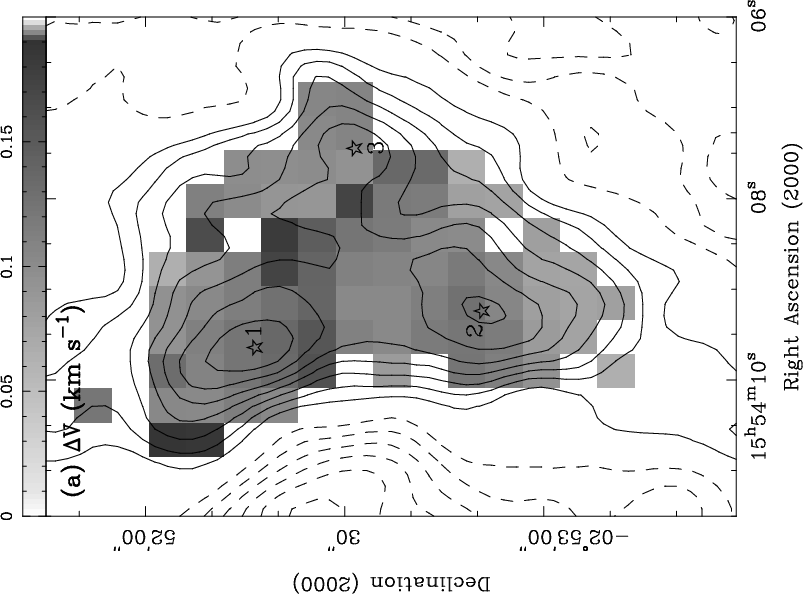}
	\includegraphics[angle=270,width=0.24\textwidth]{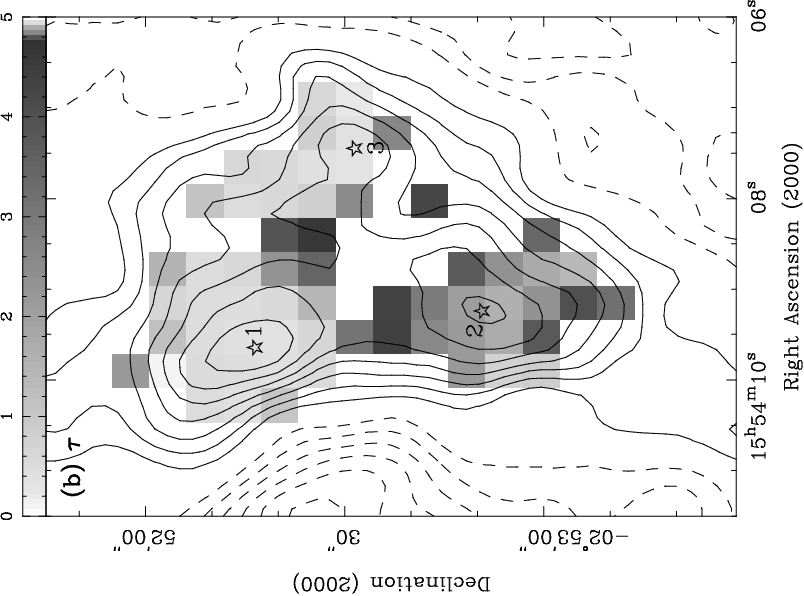}
	\includegraphics[angle=270,width=0.24\textwidth]{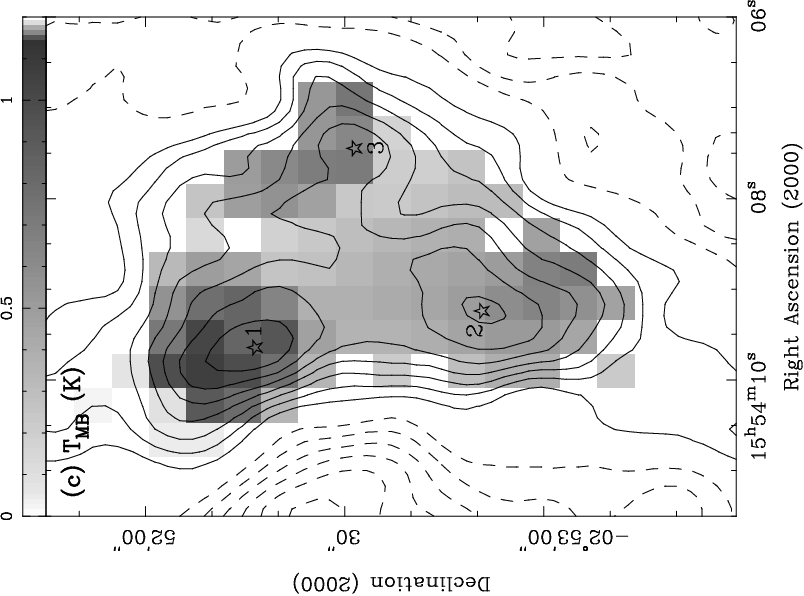} 
	\includegraphics[angle=270,width=0.24\textwidth]{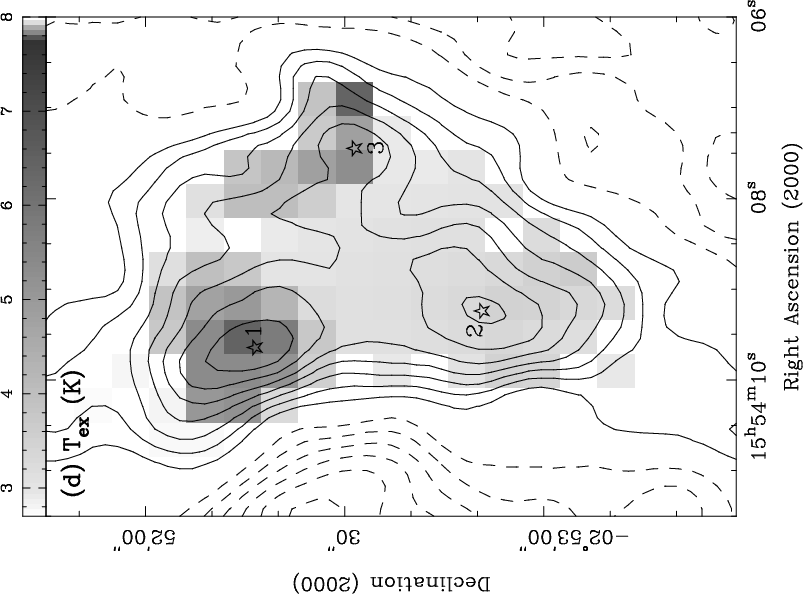}
			}
	\caption{Images of a) line width, b) optical depth, c) main beam temperature, and d) excitation temperature across the mapped area. The overlaid contours are the integrated contours from Figure \ref{fig:intmap}. The range of the displayed quantity is shown by the bars at the top of each plot. The three labeled stars show the positions listed in Table \ref{tab:hfs}. }
	\label{fig:hfs}
\end{figure*}

Figure \ref{fig:hfs} shows the remainder of the results from the hfs fitting. Panel (a) shows the linewidth variation across L183. The linewidths measured from the hfs fitting are narrow. The mean linewidth in Table \ref{tab:hfs} is only 0.119 km\,s$^{-1}$ and this is almost within the 1$\sigma$ dispersion of each measurement. The literature values for \nhp\ linewidths in L183 on peak are in the range 0.195--0.25~\kms\ \citep{2002caselli, 2005crapsi, 2007pagani}, which in turn are on the narrow side of the average linewidth of $0.250\pm0.07$~\kms\ for the 31 starless cores surveyed by \citet{2005crapsi}.

A ridge of broader linewidths of about 0.14~\kms occurs between L183-N2 and the northern part of the core containing L183-N1 and 3. The broadest linewidth along this ridge occurs between L183-N1 and L183-N3. This structure can be attributed to blending between the separate velocities of L183-N1 and 3.

A typical observed spectral line can be decomposed into non-thermal $\Delta V_{NT}$ and thermal $\Delta V_{T}$ components where the total linewidth is $\Delta V^{2} = \Delta V_{NT}^{2} + \Delta V_{T}^{2}$. The thermal line width is given by $\Delta V_{T}=c_{s} \sqrt{8\ln2}$. The sound speed is given by $c_{s}^{2} = k T_{K} / \mu m_{H}$ where $T_{K}$ is the kinetic temperature, $m_{H}$ is the mass of a hydrogen atom, $k$ is the Boltzmann constant, and $\mu$ is the mean particle mass in units of atomic mass number (for \nhp\ this is 29).

From the above we can estimate the non-thermal component from knowledge of the local sound speed, but we first need an estimate of the kinetic temperature. We have previously estimated the dust temperature in L183, using a greybody fit to the extended infra-red and submillimetre emission (in a 150-arcsec diameter aperture) centered on the dust peak, and found a value of 10K \citep{2002wak}. If this were also the kinetic temperature of the gas, then the thermal linewidth would be 0.126~\kms, and the non-thermal linewidth would be zero across the core. 

However, radiative transfer studies have shown that the extinction within pre-stellar cores can support very low temperatures at the center (as low as $\sim7$~K) with the temperature increasing radially outwards \citep{2001evans,2001zucconi,2003stamatellos}. In this scenario the dust observations give a mean temperature, but the temperature in the densest central parts, as probed by the current observations, is usually lower \citep{2003stamatellos}. 

Another way to calculate the temperature is to use the relative strength of different molecular transitions. \citet{2007daniel} and \citet{2007pagani} recently did this for higher quantum number transitions of \nhp\ and its deuterated isotopologue. The \citet{2007pagani} model found an internal temperature of 7K. The \citet{2007daniel} model found an internal temperature of 8K within a central radius of 20 arcsec. So we adopt the kinetic temperature of $T_{K}$=7~K. From this we calculate a thermal linewidth of $\Delta V_{T}$=0.105~\kms and the non-thermal linewidths listed in column 8 of Table \ref{tab:hfs}. A kinetic temperature of $T_{K}$=8~K would give a $\Delta V_{T}$=0.112~\kms.

Alternatively, we can calculate $V_{rms}$, the plane of the sky variation of $V_{lsr}$, under the assumption that it is produced by $\Delta V_{NT}$. This will be a measure of $\Delta V_{NT}$, but will not necessarily be equal to it. We calculate $V_{rms}$ by calculating the root-mean-square of the radial velocity in a $5\times5$ pixel box centered on each of the three peaks ($\sim30\times 30$ arcsec). These values are listed in column 9 of Table \ref{tab:hfs}. 

The value of $V_{rms}$ for L183-N2 is lower than for the other two sources. This might be expected from a study of Figure~\ref{fig:vlsr}, in which it can be seen that the values of $V_{lsr}$ appear fairly uniform across L183-N2. In the case of L183-N1 the mean values of $\Delta V_{NT}$ and $V_{rms}$ only differ by a few ms$^{-1}$. The variation is slightly greater for L183-N2 and 3, but for all three sources the mean of $\Delta V_{NT}$ and $V_{rms}$ is about 0.05--0.06~kms$^{-1}$. 

Particularly narrow linewidths are measured along the southern edge of the L183-N2 core. To the south of L183-N2 the total linewidth drops to $81\pm6$~m\,s$^{-1}$. If this were purely thermal it would imply a temperature of $T_{K}$=4~K. However, we note that this is the region of highest optical depth (see next section), and so we may not be sampling the whole cloud along this line of sight.


\subsection{Optical Depths}

A map of the optical depth is shown in Figure \ref{fig:hfs}b. Column 6 of Table \ref{tab:hfs} lists the total optical depth summed over the seven hyperfine lines. The optical depth for any single line is equal to the total optical depth multiplied by the fractional strength of that line. As no single line contributes more than 25\% to the total opacity, all lines \citep{2006daniel} will be optically thin for a value of $\tau<4$.

The two northern sources have total optical depths of order one-third while the southern source has an optical depth closer to two. The relatively low values of optical depth compared to earlier studies \citep[20.3,][]{2005crapsi} is due to the interferometric process resolving away the greater part of the extended material across this core. The only part that may be optically thick is the extreme southern edge.


\subsection{Temperatures}

The MIRIAD dataset was calibrated in Jy/Beam. For the CLASS hfs fitting a conversion factor of 1.07 Jy/Beam/K was used to covert the extracted spectra into units of main beam temperature $T_{MB}$. Figure \ref{fig:hfs}c shows the distribution of fitted $T_{MB}$ across the core. It is notable that the pattern of $T_{MB}$ does correlate exactly with that of the integrated intensity. The intensity valley that separates L183-N3 from L183-N1 and 2 is more pronounced. This corresponds to the northwestern edge of the emission in the 2.44~\kms channel map and the southeastern edge of emission in the 2.60~\kms channel map. 

The CLASS hfs routine returns the parameter $\chi = T_{MB}\times\tau $. Thus the quoted $T_{MB}$ error contains error contributions from the fit of $\chi$ and $\tau$. These uncertainties also reflect the difficulty in fitting the height of the line when there are only a few channels across the peak. We can calculate the excitation temperature from the equation \begin{equation} T_{MB} = \frac{\chi}{\tau} = ( T_{ex} - T_{bg} ) ( 1 - e^{-\tau} ) \end{equation} where $T_{bg}$ is the background temperature and is assumed to be equal to the cosmic microwave background (2.73~K). 

Figure \ref{fig:hfs}d shows a map of $T_{ex}$. The values for $T_{MB}$ and $T_{ex}$ at the L183-N peaks are listed in columns 7 and 10 of Table \ref{tab:hfs}. Both temperatures show a distinct increase in the region of L183-N1 where the excitation temperature is approximately equal to our assumed kinetic temperature. The CLASS hfs fitting assumes an equal excitation temperature for all hyperfine components, however, radiative transfer modeling has shown that the temperatures of the separate components can vary by up to 15\% \citep{1995caselli, 2007pagani, 2007daniel}. This effect is negligible for our purposes as we are primairly interested in the kinematics and we are in a low optical depth regime.


\subsection{Velocity gradients}
\label{rotation}

Figure \ref{fig:vlsr} shows velocity gradient vectors. These vectors are coherent across the northern half of the core -- a pattern that can be interpreted as rotation. To investigate this further we compute the the mean velocity gradient $\mathcal{G}$ over a 5 by 5 pixel box and the direction of the mean gradient for each of the L183-N sources. These parameters and their standard deviations are listed in columns 11 and 12 of Table \ref{tab:hfs}. The vectors plotted in Figure \ref{fig:vlsr} have a cut off of $\mathcal{G}>2$~\kmspc, but all non-zero values were used for the calculation of $\mathcal{G}$.

Both L183-N1 and L183-N3 have similar magnitudes and directions of velocity gradient vectors. The large standard deviation for the magnitude of the L183-N3 gradient is due to the large scatter in the gradient vectors, whereas for L183-N1 the large deviation is in part due to an increase in the gradient from the east to the west side of the core. The motions across L183-N2 show no ordered direction and the mean position angle is not statistically significant. 

\citet{2002caselli} and \citet{2005crapsi} use single dish \nhp\ observations to reconstruct the large-scale velocity field across the L183 core. \citet{2002caselli} fitted a velocity gradient of 1.19$\pm$0.08 \kmspc\ at a position angle of $-$55$\pm$3\degr, while \citet{2005crapsi} fitted a gradient of 1.4$\pm$0.1 \kmspc\ at a position angle of $-$49$\pm$1\degr. 

\citet{2007pagani} used the IRAM 30m telescope to measure \nhp\ emission at 12-arcsec intervals in an EW strip across the filament. They noted a small anti-symmetric velocity shift of a few tens of m~s$^{-1}$ at distances further than $\pm30$ arcsec from the center of the core. This gives a large-scale velocity gradient, for the cloud as a whole, of 1.2~\kmspc, which is comparable to the other single dish results quoted above.

The direction of the large scale velocity field as mapped by the single dish observations is consistent with the direction seen in our interferometric measurements. Indeed, the region where we see no gradient over L183-N2 is consistent with a region of very small vectors in figure 15 of \citet{2005crapsi}, and the vectors across the northern part of the core are also in broad agreement. 

While the direction is consistent, the magnitudes of the interferometric and single dish velocity fields appear to differ considerably. To check our gradient calculation we can work out the gradient just between L183-N2 and L183-N3 using the data in Table \ref{tab:hfs}. The velocity difference between L183-N2 and L183-N3 is 0.173 km~s$^{-1}$ and the projected separation between them is 0.018 pc. This gives a velocity gradient between them of $\sim$10 km\,s$^{-1}$\,pc$^{-1}$ which is consistent with the $\sim$8.9 km\,s$^{-1}$\,pc$^{-1}$ measured around L183-N3. 

It appears that the single dish observations are sensitive to a large-scale, low-magnitude velocity gradient while the interferometric studies are sensitive to a smaller-scale, larger-magnitude velocity gradient. Yet the direction of the gradients on the small scales is consistent with the large-scale gradients. 

The mean direction of the magnetic field in the plane of the sky observed by \citet{2004crutcher} is $28^\circ \pm 2^\circ$, which is roughly orthogonal (85$^\circ$ -- 102$^\circ$) to the velocity gradient vectors in Figure \ref{fig:vlsr} and as observed by both \citet{2002caselli} and \citet{2005crapsi}. 

\subsection{Comparison with single dish observations}

\begin{figure*}
	\centering{
	 	\includegraphics[angle=270,width=0.24\textwidth]{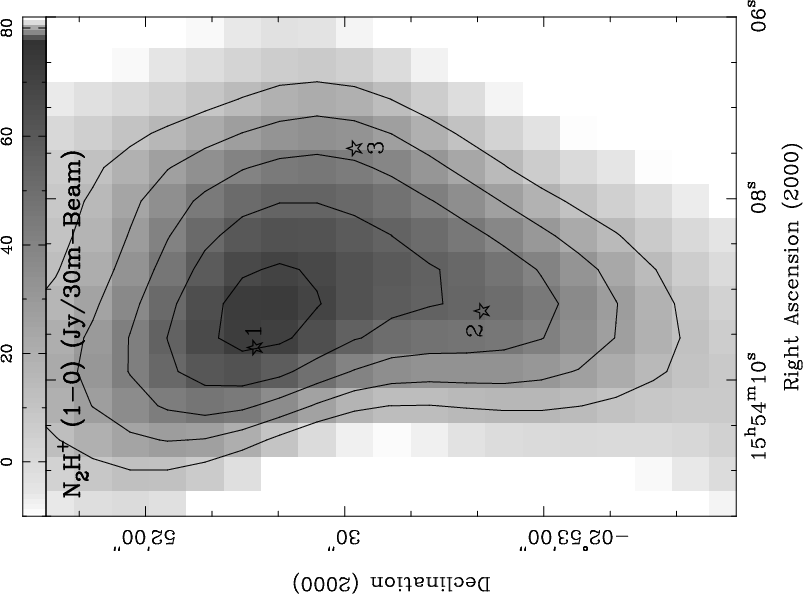}
		\includegraphics[angle=270,width=0.45\textwidth]{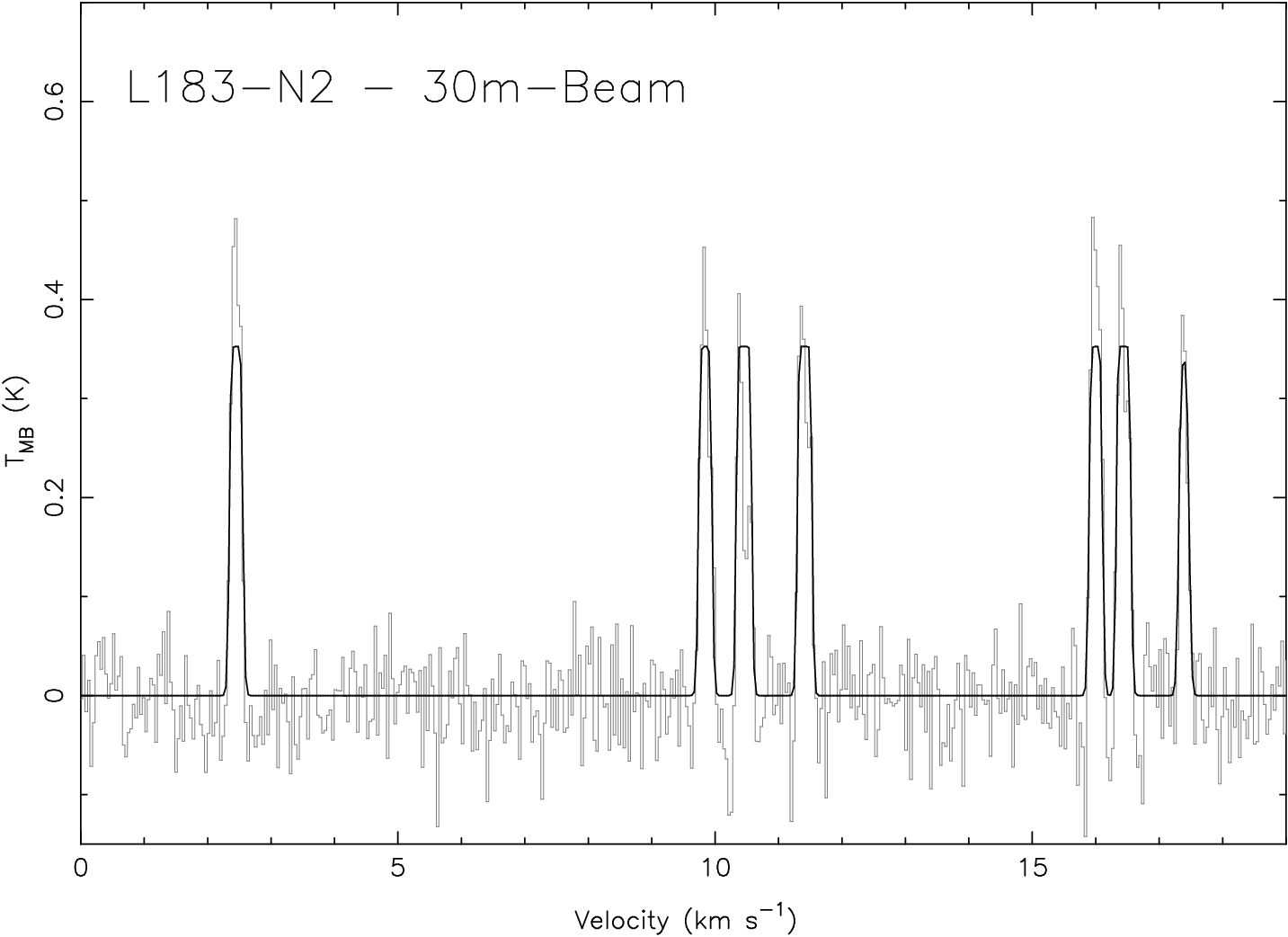}
	}
	\caption{\label{fig:30m} The BIMA interferometric \nhp\ L183 data smoothed to the equivalent resolution of a 30m single dish telescope. The left panel shows the integrated intensity with contours that start and progress in intervals of 13.8 Jy/30m-Beam. The three labeled stars show the positions listed in Table \ref{tab:hfs}. The right hand panel shows the spectrum and hfs fit for the smoothed data at the L183-N2 position. Details as per Figure \ref{fig:specs}. }
	
\end{figure*}

The left hand panel of Figure \ref{fig:30m} shows the 5.6 arcsec pixel data smoothed spatially to the resolution of a 30~m single antenna telescope. Our BIMA FWHM at 3.22~mm was $13\times10$ arcsec, the FWHM of a 30~m telescope (e.g. the IRAM dish) at the same wavelength is 27~arcsec. We use the term ``30m-Beam'' to note when we are referring to the smoothed equivalent of a 30~m single-dish beam. The smoothed map was multiplied by a factor a 5.6, the ratio of the 30m-Beam to BIMA beam integrals, in order to give a map calibrated in Jy/30m-Beam. The conversion factor between Jy/30m-Beam and K becomes 0.191~K/(Jy/30m-Beam). The smoothed map shows that the multiple peaks we detect with the BIMA interferometer would not have been resolved in {\it integrated} emission if the observations had been conducted with a 30-m diameter single dish telescope.  

While the L183-N1, 2, \& 3 substructure may not have been seen in integrated intensity maps, it could theoretically have been resolved in channel maps. However, the \citet{2007pagani} observations were along an west-east slice through L183-N2 and do not comprise a fully sampled map. Furthermore, the \citet{2000dickens} and \citet{2002caselli} \nhp\ observations of L183 were taken with the FCRAO which has an even lower resolution (50 arcsec) beam at this wavelength. Nevertheless, Figure 2 of \citet{2000dickens} shows an offset in the distribution of \nhp\ in the 2.4-2.6~\kms\ and 2.6-2.8~\kms\ channels that is comparable to us seeing higher velocity material to the west of the core. \citet{2002caselli} show a comparable vector diagram, albeit over a larger area, but they do not show channel maps. 

The (12'', 0'') position from Figure 1 of \citet{2007pagani} (15\lath 54\latm 09.3\lats\ -02\degr 52\arcmin 48\arcsec) is within the same 5.6 arcsec pixel as our L183-N2 position. We repeated  the hfs fitting for the smoothed data at that location. The results are shown in the right hand panel of Figure \ref{fig:30m}. The fitted $V_{lsr}$ for the smoothed data is $2.410\pm0.002$~\kms\ which is comparable to that fitted to the unsmoothed data. 

The linewidth $\Delta_V$ becomes $0.111\pm0.004$~\kms\ which is narrower than the value fitted to the unsmoothed data. L183-N2 is at the systemic velocity of L183, is extended spatially, and shows no velocity structure. Therefore spatially smoothing the wings of the emission (e.g., channels 2.29 and 2.52\kms\ in Figure \ref{fig:linegrid}) would move flux out of the beam whereas smoothing the extended emission in channels near the line center (e.g., channels 2.40 and 2.44\kms\ in Figure \ref{fig:linegrid}) would bring more flux into the beam. This differential effect will naturally cause the linewidth to narrow.The opacity becomes $\tau=3.2\pm0.6$, which is higher than for the unsmoothed data, but is still substantially lower than reported for single-dish observations. 

The line temperature for our smoothed L183-N2 position was fitted as $T_{MB}=0.4\pm0.1$ which agrees with our fit to the unsmoothed data at that position. However, inspection of Figure \ref{fig:30m} shows that the fit has been unable to recover the full extent of the line temperature. \citet{2007daniel} and \citet{2007pagani} respectively published spectra towards L183 with $T_{A}^{*}$ and $T_{R}^{*}$ values of 1.5-2.0~K. The (12'', 0'') position of \citet{2007pagani} had a peak  $T_{R}^{*}\sim1.8$~K. Even allowing for the BIMA main beam efficiency \cite[83\%, ][]{1995lugten} our line temperatures are only about a quarter of those found in the single dish observations. This agrees with the result of our optical depth fitting to the unsmoothed data. A substantial fraction, possibly 75\%, of the extended \nhp\ emission from L183 is being resolved away by the interferometric process. It is this process that has revealed the clumps buried within the extended L183 core.


\section{Discussion}
\label{discuss}

We have studied the pre-stellar core L183 with high-resolution BIMA interferometric observations, and have seen that the core breaks up into three separate clumps, which we have labeled L183-N1, 2 \& 3. The clump L183-N2 is roughly coincident with the densest central part of L183. Furthermore, the velocity of L183-N2 is exactly that of the systemic velocity of the whole of L183. We therefore deduce that L183-N2 is at the center of L183, and that the L183 core has formed two further fragments, L183-N1~\&~3.

The virial mass for a uniform density, turbulent, rotating molecular cloud threaded by a uniform magnetic field can be written as $M_{vir}=M_{\Delta V}+M_{B}+M_{\mathcal{G}}$ where the three terms on the right hand side represent the turbulent and thermal virial mass, the magnetic critical mass, and the kinematic mass of the system from Kepler's third law. This can be written as 
\begin{equation}
	M_{vir} = 210\Delta V^{2} r + 9.3Br^{2} + 233\mathcal{G}^{2}r^{3} \mathrm{\ M}_{\odot}
\end{equation}
where $\Delta V$ is the linewidth in \kms, $r$ is the radius of the cloud in parsec, $B$ is the magnetic field strength in $\mu$G, and $\mathcal{G}$ is the velocity gradient due to rotation in \kmspc. Here we have used the expression for $M_{\Delta V}$ from  \citet{1988maclaren}. \citet{2004crutcher} used the Chandrasekhar-Fermi method to estimate a field strength of $B=80$~$\mu$G towards L183. The approximate radius of L183-N1 and L183-N2 is $r\sim0.01$~pc and the radius of L183-N3 is $r\sim0.005$~pc. Based on these values and the parameters listed in Table \ref{tab:hfs} we estimate $M_{vir}\sim0.1$~M$_{\odot}$ for L183-N1 and 2 and $M_{vir}\sim0.03$~M$_{\odot}$ for L183-N3. 

For a virialised system the virial mass $M_{vir}$ will be balanced by the actual mass of the fragment $M_{clump}$ plus a surface term $M_{P}$ due to a confining medium. The material we have resolved away will act as this surface pressure term. By equating an expression for the energy in a confining isothermal pressure to the standard formula for gravitational potential energy it can be shown that $M_{P}=316r^{2}\sqrt{n_{4}T_{10}}$ where $n_{4}$ is the number density of the confining medium in units of $10^{4}$~cm$^{-2}$ and $T_{10}$ is the gas temperature in units of 10~K. Using $n_4=100$ from \citet{2005kwa} and our assumed gas temperature of 7K we estimate $M_{P}\sim0.3$~M$_{\odot}$ for L183-N1 and 2 and $M_{P}\sim0.07$~M$_{\odot}$ for L183-N3. In each case the value of $M_{P}$ is 2-3 times greater than the value of $M_{vir}$. Even without estimating the actual mass of these fragments, we can see that they are each pressure confined and will probably each collapse to form a protostar. 

{\it Spitzer} archival data of L183 was checked for the presence of a young stellar object (YSO). No YSO candidates were identified towards L183 based on the selection criteria \citep{2007harvey,2007rebull,2008harvey} used by the {\it Spitzer} c2d \citep{2003evans} and Gould Belt \citep{2008allen} legacy surveys. There are no 2MASS sources within the L183 fields shown in this paper. We therefore confirm that L183 and each of the L183-N clumps are pre-stellar in nature and do not contain embedded YSOs.

The large-scale structure of L183 is elongated approximately north-south at a position angle of $\sim-20^\circ$. \citet{2004crutcher} gave the position angle of the long-axis of the SCUBA emission as $\sim-15^\circ$. These position angles are consistent with the elongation orientation of L183-N2. We have examined the detailed velocity structure of the core and found that L183-N1 \& 3 each have velocities slightly offset from that of L183-N2 -- one positive and one negative. Furthermore, there is a velocity gradient in the general direction from L183-N1 to L183-N3 of $\sim$6 -- 9 km\,s$^{-1}$\,pc$^{-1}$, over a separation of $\sim$0.016~pc (see Table \ref{tab:hfs}). We refer to this hereafter as the small-scale velocity gradient. 

The mean direction of the small-scale velocity gradient across the entire core is $-60\pm62^\circ$. However, the mean direction for ``high-velocity vectors'' (defined as those vectors with $V_{lsr}$ greater than 0.1~\kms\ away from the systemic velocity, i.e. $|V_{lsr}-2.415|>0.1$~\kms) is $-72\pm2^\circ$. The high-velocity vectors have a more coherent direction than vectors with a velocity close to the systemic velocity. The mean direction for the subset of high-velocity vectors in L183-N1 is $-68\pm30^\circ$ (9 vectors), and for L183-N3 is $-79\pm5^\circ$ (5 vectors). These directions are measured over the same areas used in Section \ref{rotation}. 

Therefore the axis of rotation for the small-scale gradient is $+22\pm30^\circ$ for L183-N1 and $+11\pm5^\circ$ for L183-N3, assuming that the axis is orthogonal to the direction of the velocity gradient. We note that the axis of rotation given by the mean of all 14 high-velocity vectors in L183-N1 and N3 is $+18\pm19^\circ$. This is approximately parallel to the direction of the magnetic field across L183 \citep[$+28\pm2^\circ$,][]{2004crutcher}, but is at an angle of $\sim33^\circ$ to the core long axis. This is indicating that perhaps the magnetic field may be playing an important role in the evolution of L183. It also indicates that, for L183, the axis of rotation shares the same offset from the orientation of the cloud as is found for the magnetic fields in other prestellar cores \citep{2004crutcher,2006kirk,2009wskn}.

We therefore interpret our data of L183 as a rotating elongated filamentary core, whose rotation axis roughly coincides with its magnetic field direction. L183-N2 is on-axis, and therefore essentially stationary. L183-N1 \& 3 lie either side of the rotation axis and have similar magnitude line-of-sight velocities in opposite directions. The small-scale velocity gradient therefore tells us about the rotation near the center of L183.

On the larger scale of the L183 pre-stellar core as a whole, previous single-dish work by \citet{2005crapsi} has found a core 0.09 $\times$ 0.05 pc in extent -- see their figure 15. These authors also saw a velocity gradient roughly parallel to the gradient we observed, but on a larger spatial scale. They see a gradient of $\sim$1.4 km\,s$^{-1}$\,pc$^{-1}$, over a scale of $\sim$0.05~pc. We refer to this hereafter as the large-scale velocity gradient.

The ratio of the spatial scales of the large-scale and small-scale velocity gradients is 0.05/0.016 $\sim$ 3. The ratio of the velocity gradients themselves is (6 -- 9)/1.4 $\sim$ 4 -- 6. Given the uncertainties in all of the foregoing estimates and calculations, the agreement of these two ratios to within a factor of two is very interesting. This is exactly what would be predicted by the law of conservation of angular momentum if L183 were collapsing and spinning up as it collapses.

We therefore hypothesize that all of our data of L183 can be explained in terms of a collapsing, fragmenting elongated core, that is rotating about the magnetic field direction and spinning up as it collapses. We note finally that our hypothesis is consistent with the single-dish data of both \citet{2005crapsi} and \citet{2007pagani}, whose large-scale velocity vectors also appear to increase in magnitude away from the center of L183. Thus our hypothesis is also consistent with previous data of L183.

\section{Conclusions}

We have used BIMA to make deep \nhp\ (1-0) maps of the pre-stellar core L183. The interferometric process resolved away the majority of the extended \nhp\ emission and revealed three clumps which we have labeled L183-N1, 2 \& 3. L183-N2 is at the established systemic velocity of L183 and is coincident with the position of the peak of the dust emission. We detect a velocity gradient across the core. The direction of the gradient is consistent with previously published lower-resolution studies, but its magnitude is larger than previously seen. We interpret this pattern as the spin up and fragmentation of a rotating and collapsing pre-stellar core.


\section{Acknowledgments}

We thank the anonymous referee for providing constructive comments that have improved the contents of this paper. Work for this paper was conducted in part with NSF post-doctoral support at the University of Illinois from grant NSF AST 02-28953, and in part with the support of the UK Science and Technology Facilities Council (STFC) via the Cardiff Astronomy Rolling Grant. BIMA was operated by the Berkeley-Illinois-Maryland Association with NSF and University Funding. The BIMA facilities have since been merged with the OVRO array to form the present CARMA array supported at the University of Illinois by University funding and by grant NSF AST 05-40459.

{\it Facilities:} \facility{BIMA ()}

\bibliographystyle{apj}

\end{document}